\documentclass{aa}
\usepackage{graphicx}
 
\begin{document}
 
\title{The K20 survey. V The evolution of the near-IR Luminosity Function 
	\thanks{Based on observations made at the European
       Southern Observatory, Chile (ESO LP 164.O-0560).}
        }


\author{L. Pozzetti \inst{1}
        \and
        A. Cimatti \inst{2}
        \and
        G. Zamorani \inst{1}
        \and
        E. Daddi \inst{3}
	\and
        N. Menci \inst{4}
        \and
        A. Fontana \inst{4}
        \and
        A. Renzini \inst{3}
        \and
	M. Mignoli \inst{1}
        \and
        F. Poli \inst{5}
        \and
        P. Saracco \inst{6}
	\and
        T. Broadhurst \inst{3,7}
        \and
        S. Cristiani \inst{8,9}
        \and
        S. D'Odorico \inst{3}
        \and
        E. Giallongo \inst{4}
        \and
        R. Gilmozzi \inst{3}
        }

\offprints{Lucia Pozzetti, \email{lucia@bo.astro.it}}

\institute{
Istituto Nazionale di Astrofisica,
Osservatorio Astronomico di Bologna, Via Ranzani 1, I-40127 Bologna, Italy
\and  
Istituto Nazionale di Astrofisica,
Osservatorio Astrofisico di Arcetri, Largo E. Fermi 5, I-50125 Firenze, Italy
\and 
European Southern Observatory, Karl-Schwarzschild-Str. 2, D-85748, Garching, 
Germany
\and 
Istituto Nazionale di Astrofisica,
Osservatorio Astronomico di Roma, via Dell'Osservatorio 2, Monteporzio, Italy
\and 
Dipartimento di Astronomia, Universit\`a ``La Sapienza'', Roma, Italy
\and 
Istituto Nazionale di Astrofisica,
Osservatorio Astronomico di Brera, via E. Bianchi 46, Merate, Italy
\and 
Racah Institute for Physics, The Hebrew University, Jerusalem, 91904, Israel
\and 
ST, European Coordinating Facility, Karl-Schwarzschild-Str. 2, D-85748, 
Garching, Germany
\and 
Istituto Nazionale di Astrofisica,
Osservatorio Astronomico di Trieste, via Tiepolo 11, Trieste, Italy
}

\date{Received ... ; Accepted ...}

\abstract{
We present the galaxy rest-frame near-IR Luminosity Function (LF) and 
its cosmic evolution to $z\sim1.5$ based on 
a spectroscopic survey of a magnitude limited sample of galaxies 
with $K_s<20$ (the K20 survey, Cimatti et al. 2002b). 
The LFs have been derived in the rest-frame $J$ and $K_s$ bands. Their
evolution is traced using three different redshift bins ($z_{mean}\simeq
0.5, 1, 1.5$) and comparing them to the Local near-IR Luminosity Function. 
The luminosity functions at different redshifts are fairly well fitted by 
Schechter functions at $z<1.3$. 
The faint-end of the LFs ($L<L^*$) is consistent with the
local estimates, with no evidence for a change either in the slope or
normalization up to $z<1.3$.
At higher redshift this part of the luminosity function is not well
sampled by our data.
Viceversa, the density of luminous galaxies 
($M_{K_s}-5$ log$ h_{70}<-25.5$) is higher than locally
at all redshifts and relatively constant or mildly increasing with redshift 
within our sample.
The data are consistent with a {\it mild luminosity 
evolution} both in the $J$- and $K_s$-band up to $z\simeq1.5$, with an
amplitude of about $\Delta M_J \simeq -0.69\pm0.12$ and 
$\Delta M_K \simeq -0.54\pm0.12$
at $z\sim1$. Pure density evolution is not consistent with the 
observed LF at $z\le1$. 
Moreover, we find that {\it red and early-type galaxies 
dominate the bright-end of the LF}, 
and that their number density shows at most a 
small decrease ($<30\%$) up to $z\simeq 1$, 
thus suggesting that massive elliptical galaxies were already in place 
at $z\simeq 1$ and {\it they
should have formed their stars and assembled their mass at higher
redshift.} 
There appears to be a correlation of the optical/near-IR colors 
with near-IR luminosities, the most luminous/massive
galaxies being red/old, the low-luminous galaxies being instead dominated
by blue young stellar populations. 
We also investigate the evolution of 
the near-IR comoving luminosity density to $z\simeq1.5$, finding a slow 
evolution with redshift ($\rho_\lambda(z)= \rho_\lambda(z=0)
(1+z)^{\beta(\lambda)}$ with $\beta(J)\simeq0.70$ and
$\beta(K_s)\simeq0.37$). 
Finally, we compare the observed
LFs with the predictions of a set of the most updated hierarchical merging 
models. Such a comparison shows that the current versions of hierarchical 
models 
overpredict significantly the density of low luminosity galaxies at 
$z\le1$ and underpredict the density of luminous galaxies at $z\ge1$,
whereas passive evolution models are more consistent with the 
data up to $z\sim1.5$.  
The GIF model (Kaufmann et al. 1999) shows 
a {\it clear deficiency of red luminous galaxies at $z\sim1$} compared to our
observations and predicts a
decrease of luminous galaxies with redshift not observed in our sample. 
}

\authorrunning{Pozzetti and K20 collaboration}
\titlerunning{The evolution of the near-IR LF} 

\maketitle
 

\section{Introduction}\label{sec:intro}

Over the past few years, a wealth of observations from deep surveys of 
optically-selected high-redshift galaxies (e.g. Madau et al. 1996, 
Steidel et al. 1999), complemented by 
observations in the far-IR/sub-mm (Hughes et al. 1998, Barger et al.
1999), allowed significant
progress in our understanding of the evolution of galaxies from the 
present-epoch back to $z\simeq 4$ and beyond 
(Steidel et al 1999; Madau, Pozzetti, \& Dickinson 1998). 
However, since these samples of high-$z$ galaxies were all selected at 
optical or far-IR/sub-mm wavelengths, they are dominated by objects with 
on-going star formation. Therefore, such studies placed constraints more on 
the evolution of the star birth rate activity than on the formation and 
assembly of stellar systems through cosmic time.

The study of faint galaxy samples selected in the near-infrared 
represents an important and complementary possibility to address 
the still open questions on how the formation and evolution of 
massive systems evolved with time compared to the predictions of
the different theoretical scenarios (Broadhurst et al. 1992).
Another advantage of the near-IR selection (in particular in the
$K$-band) is that the k--corrections are relatively insensitive 
to galaxy type and fairly small also at high redshift (Cowie et 
al. 1994), and the dust extinction effects are less severe than in 
optical samples. 

Since the rest-frame near-IR light is a relatively good tracer of the galaxy 
stellar 
mass (Gavazzi et al. 1996; Madau, Pozzetti \& Dickinson 1998) the 
near-IR galaxy Luminosity Function (LF)  can
provide a reasonable estimate of the Galaxy
Stellar Mass Function (GSMF). Only recently the 2MASS (Jarrett et al. 2000) 
surveys allowed accurate determinations of the local near-IR luminosity
and of the Galaxy Stellar-mass functions (Cole et al. 2001, Kochanek et al.
2001), while only few 
attempts have been made to reconstruct their evolution with redshift 
using deep surveys of near-IR selected samples (i.e. Cowie et al. 1996; 
Cohen et al. 1999, Cohen 2002).

In order to address the above questions, we performed a new
spectroscopic survey of a complete sample of galaxies selected
with $K_s<20$ (the K20 survey; {\tt http://www.arcetri.astro.it/$\sim$
k20/}). The survey and the sample are described
in detail in Cimatti et al. (2002b, hereafter Paper III), while the
spectral and clustering properties of the 
Extremely Red Objects (EROs) are discussed in Cimatti et al. (2002a, Paper I) 
and Daddi et al. (2002, Paper II), respectively.
The redshift distribution for the whole sample of $K_s$-band selected 
galaxies is given in Cimatti et al. (2002c, Paper IV).
Here we recall that the
K20 sample includes 546 objects to $K_s<20$ (Vega system),
selected from a 32.2 
arcmin$^2$ area of the Chandra Deep Field South (CDFS; Giacconi et al.
2001) and from a 19.8 arcmin$^{2}$ field centered at 0055-269.  
The total area of the two fields is 52 arcmin$^2$.
Optical multi-object spectroscopy was mainly obtained with the ESO VLT + 
FORS1 and FORS2, while a small fraction ($\sim 4\%$) of the objects
was observed with near-IR spectroscopy using the VLT + ISAAC.
We have imaged the 0055-269
field over 10 bands ($UBGVR R_w IzJK_s$), obtained with the ESO
NTT + SUSI2 ($UBGV R_w I$) and SOFI ($JK_s$), and VLT + FORS1 ($R$ and
$z$), for a total of about 45 hours
of integration. In the CDFS field, we used a combination of
public EIS NTT data ($UK_s$) and deep FORS1 images ($BV R Iz$,
courtesy of P. Rosati \& M. Nonino).
The spectroscopic redshift completeness is 
94\% and 87\% for $K_s<19$ and $K_s<20$, respectively, and it increases
to 98\% if we include the photometric redshifts obtained with the 
deep multi-band imaging for the spectroscopically unidentified or unobserved
objects (Cimatti et al. 2002b). 
The K20 sample is the largest and most complete spectroscopic
sample of galaxies with $K_s<20$ available to date.

\begin{figure}[ht]
\resizebox{\hsize}{!}{\includegraphics{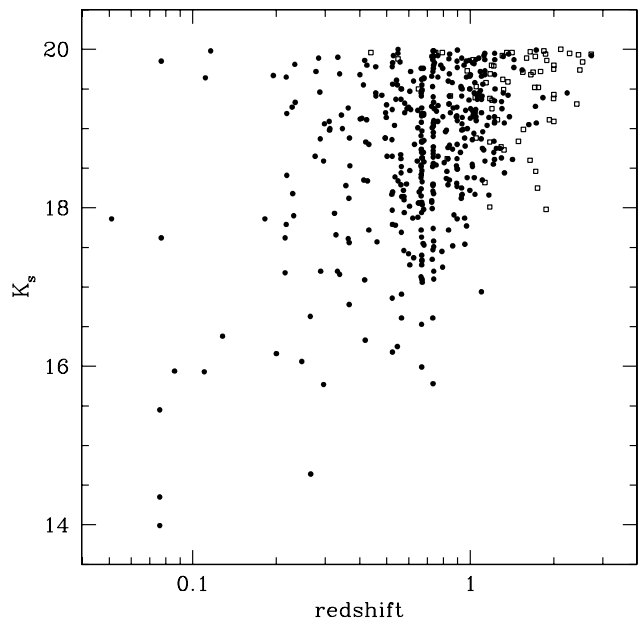}}
\caption{\footnotesize  
The magnitude-redshift diagram for all the galaxies in the
spectroscopic sample to $K_s<20$.
Filled circles represent galaxies identified spectroscopically, while 
empty squares are unidentified or unobserved galaxies plotted at $z=z_{phot}$.
}
\label{fig:Kz}
\end{figure}

\begin{figure}[h]
\resizebox{\hsize}{!}{\includegraphics{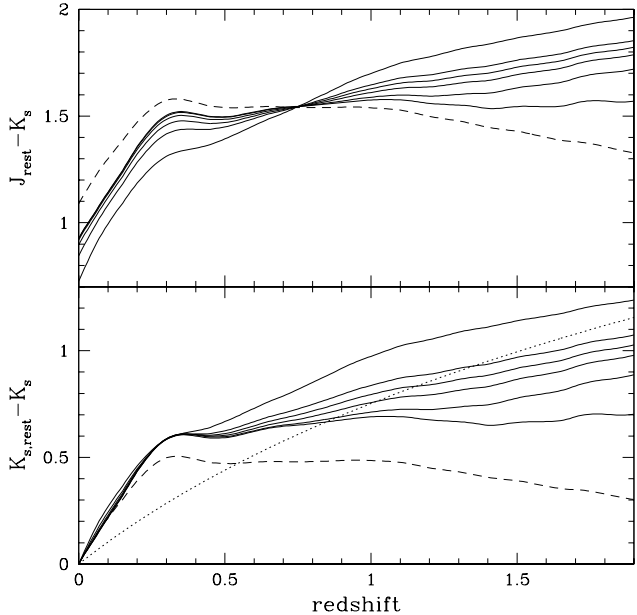}}
\caption{\footnotesize  
K-corrections colors (see text) as a function of redshift:
$J_{rest}-K_s$ (top panel) and $K_{s,rest}-K_s$ (bottom panel).
Different curves are derived from different
spectral models which at low redshift reproduce spectra and colors
of local E, S0, Sa, Sb, Sc, Im (solid lines from bottom to top at $z>1$) 
and dusty star-forming galaxies (dashed line; see text). Dotted line shows 
the $2.5 log (1+z)$ term for comparison.
}
\label{fig:kcorr}
\end{figure}

In this paper, we investigate the evolution of the near-IR 
luminosity function up to $z\simeq1.5$ based on the K20 survey sample.
Thanks to the higher statistical significance and completeness of our
sample, it is possible to use the LF evolution to place new and
more stringent constraints on the formation and evolution of massive
galaxies than was possible from previous surveys
(Cowie et al. 1996, Cohen et al. 1999, Cohen 2002). 
In particular our survey has the advantage to include a complete 
sample of EROs, partially with spectroscopic identifications, which usually 
were not included in previous spectroscopic surveys (e.g. Cohen 2002) because 
of their selection.
Finally, we compare 
the LF evolution with the predictions of two competitive scenarios of galaxy 
evolution: the Pure Luminosity Evolution
(PLE) and the Hierarchical Merging Models (HMM). We adopt $H_0=70$ km 
s$^{-1}$ Mpc$^{-1}$, $\Omega_m=0.3, \Omega_\Lambda=0.7$.


\section{The K20 spectroscopic sample}\label{sec:sample}

\begin{figure}[ht] 
\resizebox{\hsize}{!}{\includegraphics{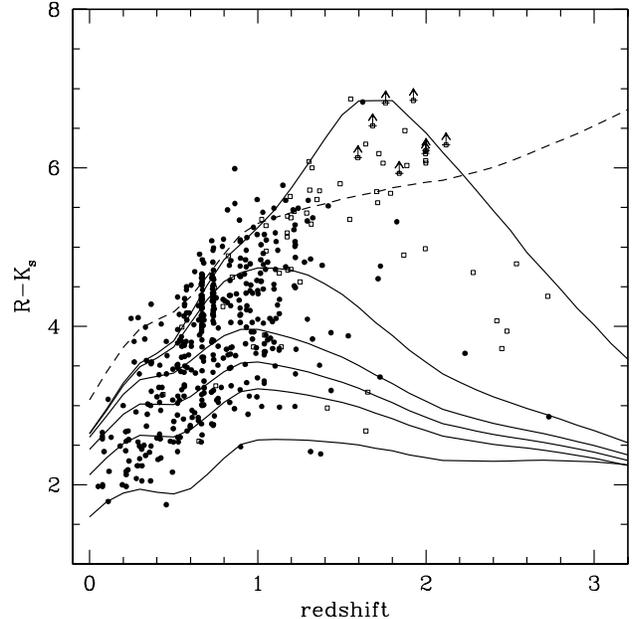}}
\caption{\footnotesize  
The color-redshift ($R-K_s$ vs. $z$) diagram for galaxies in the
spectroscopic sample to $K_s<20$.
Different symbols as in Fig. \ref{fig:Kz}.
Arrows indicate $3 \sigma$ lower limits in $R-K_s$ colors
for objects not detected in the $R$ band.
The curves show different spectral models as in Fig. \ref{fig:kcorr}.
}
\label{fig:RKz}
\end{figure}

From the total K20 sample of $546$ objects with $K_s\leq20$ we have extracted 
a sample of $489$ galaxies, after excluding objects classified as stars and 
AGN on the basis of their spectra.  
In addition to spectroscopic redshifts, photometric redshifts have been
derived (Cimatti et al. 2002b) with very high accuracy 
($\langle z_{spe}- z_{phot} \rangle=\langle \Delta z \rangle= 0.012$ 
and $\sigma_{\Delta z}=0.089(1+z_{spe})$).
Such an accuracy, made possible by the numerous photometric bands
and
the precision of the magnitude measurements,
is consistent with previous results on the HDF-N data set (Fontana et al 2000)
and with simulations made by Bolzonella et al. (2000)
at high redshift ($1<z<2$), where our sample is the largest to date. 
The magnitude--redshift distribution of selected galaxies
is shown in Fig. \ref{fig:Kz}. Thanks to the depth and completeness
of our sample, the coverage of the $K_s$ -- redshift plane is
such that the near-IR LF can be derived with good accuracy
and over a relatively large range of magnitudes at least up
to $z\sim1$, still sampling with relatively good statistics
the high luminosity part of the LF up to $z\sim2$.
Since the sample spans a wide range in luminosity, redshift
and look-back time, 
it is therefore well suited to study the evolution of the 
near-IR Luminosity Function within the sample itself and in comparison to the 
local population.  


\section{The estimate of the rest-frame near-IR luminosities} \label{sec:LIR}

The small variance of k-corrections in the $K_s$-band (Cowie et al. 
1994, Mannucci et al. 2001) greatly simplifies the analysis with respect
to optically-selected samples of galaxies. Moreover, we take advantage
of the fact that at $z\simeq0.8$ the observed $K_s$-band corresponds 
to rest-frame $J$-band.
In order to compute the near-IR luminosity function, we derive 
the rest-frame $J$- and 
$K_s$-band absolute magnitudes ($M_J$ and $M_{K_s}$) according
to the following relations:

\begin{equation}\label{eq:MJ}
M_J=K_s-5 log (d_L(z)/10 pc) + (J_{rest}-K_s)(z)
\end{equation}
\begin{equation}\label{eq:MK}
M_{K_s}=K_s-5 log (d_L(z)/10 pc) + (K_{s, rest}-K_s)(z),
\end{equation}
where $K_s$ is the total apparent magnitude measured with SExtractor
(see Cimatti et al. 2002b, Paper III, for more details), $d_L(z)$ is the 
luminosity 
distance at redshift $z$ and ($J_{rest}-K_s$) and ($K_{s,rest}-K_s$) are 
the ``k-correction colors", i.e. the difference between rest frame and 
observed magnitude which includes also the $2.5 log (1+z)$ term (see also Lilly 
et al. 1995). It should be noted that in Eq. (\ref{eq:MJ}) the last term can be
written as a conventional k-correction for the $K_s$-band (as in Eq.
\ref{eq:MK}) plus a rest-frame (J-$K_s$) color. 
The k-correction colors as a function of redshift are
plotted in Fig. \ref{fig:kcorr}. This figure clearly shows the small
variation of the k-corrections for different spectral types
over most of the redshift range covered by our data
(for a comparison with a similar plot in the optical bands see
Lilly et al. 1995).

\begin{figure}[ht] 
\resizebox{\hsize}{!}{\includegraphics{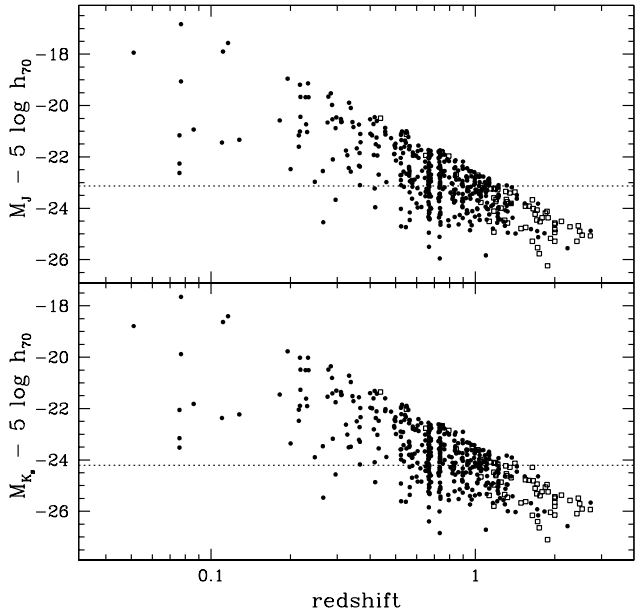}}
\caption{\footnotesize  
The absolute $J$ and $K_s$ magnitudes as a function of redshift 
derived (see text) for all galaxies in the spectroscopic sample 
to $K_s<20$. Symbols as in Fig. \ref{fig:Kz}. Local estimates of
$M^*_J$ and $M^*_{K_s}$  (dotted horizontal lines) 
from Cole et al. (2001). 
}
\label{fig:Mz}
\end{figure}

These k-corrections for different spectral types are computed
using the Bruzual \& Charlot  (1993) models (GISSEL 2000 version), 
in order to reproduce 
spectra, and k-corrections of local galaxies (E, S0, Sa, Sb, Sc, Im) 
at low redshift 
(cf. Pozzetti et al. 1996, Mannucci et al. 2001) and the color-redshift diagram 
($R-K_s$ vs. $z$) observed in our sample (Fig. \ref{fig:RKz}). 
In order to take
into account the dusty star-forming population, we introduce an additional 
spectral type dominated by star formation and strong dust extinction
consistently describing the SED of red emission line 
objects and reproducing the sub-class of emission line EROs with
$R-K_s>5$ (cf. Paper I). 
We have adopted an exponentially declining star formation history 
($SFR\propto exp(-t/\tau)$), with $\tau=0.3,1,2,4,\infty$ Gyrs
(E, S0, Sa, Sb, Sc galaxies, respectively), 
the Salpeter (1955) Initial Mass Function (IMF) and an age of 12.5 Gyrs at $z=0$
(i.e. a formation redshift $z_f=5.7$). A model with an age of 1 Gyr
at all redshifts is adopted for Im galaxies,
while  a constant SFR and $E(B-V)=0.5$ has been assumed for the dusty 
star-forming galaxies (cf. Paper I).  
We have then assigned a spectral model 
to each galaxy according to their $R-K_s$ color and spectral type. 

We tested that different assumptions on the k-corrections, i.e. by varying 
model parameters or using the fitting of multi-band photometry when 
available, did not affect significantly our results. 
For example, the k-correction estimates which make use of multi-band 
fitting of the SED are in excellent agreement with our estimates 
from $R-K_s$ colors alone, with no systematic shifts in magnitude and a 
dispersion 
of about 0.08 in the $J$-band and 0.13 in the K-band up to $z\sim1.5$.
Varying parameters in the synthetic spectral models, i.e. the
IMF (Salpeter 1955, Scalo 1986 and Kennicutt 1983) and the 
formation redshift ($z_f=2,3,6$), 
the systematic shifts and dispersions at $z<2$ are always less than 0.05 
magnitudes. We note that magnitudes derived using the 
k-corrections estimated from observed spectra of local galaxies 
(Mannucci et al. 2001, Cowie et al. 1994) would be brighter by
about 0.1 magnitudes at $z>0.7$, and could be considered upper limits due to 
the possible evolutionary effect on the galaxy spectrum. 
We are therefore confident that the derived luminosities 
are robust with respect to varying this set of assumptions. 
The resulting effects on the Luminosity Functions induced by changes in
our assumptions are all
comparable with or smaller than the statistical errors (see Sect.
\ref{sec:LFz}).

Fig. \ref{fig:Mz} shows the absolute magnitudes of all galaxies
in the spectroscopic sample to $K_s<20$ as derived according
to our adopted recipes. For the unidentified or unobserved 
galaxies we have adopted the photometric redshift $z_{phot}$. 
The photometric threshold of the survey, $K_s<20$, is clearly visible in the 
data and translates into different luminosity limits as a function of redshift, 
varying from less 
than $\sim 0.1 L^*$ at $z<0.5$ to about $\sim 0.4 L^*$ at $z\simeq 1$, 
assuming $M^*_J=-23.13+5$log$h_{70}$ and 
$M^*_{K_s}=-24.21+5$log$h_{70}$ 
(horizontal lines in Fig. \ref{fig:Mz}) 
for the local LF (Cole et al. 2001). 
Adopting the mean stellar mass-to-light ratio and representative stellar 
mass (${\cal M}_{stars}/L_K=1.32$ in solar unit and 
${\cal M}^*_{stars}=1.44\times 10^{11} 
h_{70}^{-2} M_\odot$ for the Salpeter IMF) in the local 
universe (Cole et al. 2001), the 
limits in luminosities correspond to $\sim 0.1$--$0.5 {\cal M}^*_{stars}$ or 
even smaller stellar masses since the ${\cal M}_{stars}/L_K$ increases as 
the stellar population ages (Madau, Pozzetti, \& Dickinson 1998).

\begin{table*}[t]
\begin{flushleft}
\caption{Schechter Luminosity Function parameters
($\Phi(M)=0.4ln(10)\Phi^*10^{-0.4(M-M^*)(\alpha+1)}exp[-10^{-0.4(M-M^*)}]$)}
\protect\label{tab:LFP}
\begin{tabular}{cccccc}
\noalign{\smallskip}
\hline
\noalign{\smallskip}
\multicolumn{1}{c}{Band} & 
\multicolumn{1}{c}{$z$ range } & 
\multicolumn{1}{c}{$M-5$ log $h_{70}$ range} & 
\multicolumn{1}{c}{$\alpha$} & 
\multicolumn{1}{c}{$M^*-5$ log $h_{70}$} & 
\multicolumn{1}{c}{$\Phi^* h_{70}^{-3}$ ($10^{-3}$ Mpc$^{-3}$)} \\ 
\noalign{\smallskip}
\noalign{\smallskip}

$J$ & $0.20-0.65$ & $[-18.7, -25.2]$ & $-1.22^{+0.22}_{-0.20}$ & $-23.89^{+0.51}_{-0.69}$ & $1.99^{+1.4}_{-1.1}$ \\ 
\noalign{\smallskip}
$J$ & $0.75-1.30$ & $[-21.8, -26.3]$ & $-0.86^{+0.46}_{-0.44}$ & $-23.75^{+0.40}_{-0.51}$ & $3.44^{+1.1}_{-1.5}$ \\ 
\noalign{\smallskip}
$J$ & $1.30-1.90$ & $[-23.3, -26.3]$ & $-$ & $-$ & $-$ \\ 

\noalign{\medskip}

$K_s$ & $0.20-0.65$ & $[-19.5, -26.0]$ & $-1.25^{+0.25}_{-0.20}$ & $-24.87^{+0.54}_{-0.73}$ & $1.78^{+1.5}_{-0.9}$ \\ 
\noalign{\smallskip}
$K_s$ & $0.75-1.30$ & $[-22.7, -27.2]$ & $-0.98^{+0.47}_{-0.42}$ & $-24.77^{+0.42}_{-0.55}$ & $2.91^{+1.3}_{-1.4}$ \\ 
\noalign{\smallskip}
$K_s$ & $1.30-1.90$ & $[-24.1, -27.1]$ & $-$ & $-$ & $-$ \\ 
\noalign{\smallskip}
\hline
\end{tabular}
\end{flushleft}
\end{table*}

\section{The estimate of the near-IR Luminosity Function}
\label{sec:LF}

After the estimate of the rest-frame near-IR luminosities, we 
compute the $J$- and $K_s$-band Luminosity Functions in the 
following redshift bins: 
(a)  $0.2<z<0.65$ ($z_{mean}\simeq0.5$),
(b)  $0.75<z<1.3$ ($z_{mean}\simeq1.0$) and
(c)  $1.3<z<1.9$ ($z_{mean}\simeq1.5$). 
The redshift bin 0.65$<$z$<$0.75 has been excluded from this
analysis because dominated 
by two groups/clusters (see Fig. 1 and Paper III for details).  
The first two redshift bins include respectively 132 and 170 galaxies,
of which only few objects (2 and 21 respectively) do not have a
spectroscopic redshift. The highest redshift bin includes a smaller number of 
galaxies (42) with a high fraction of photometric redshifts (60\%).

The Luminosity Functions in each redshift bin are estimated using both 
the $1/V_{max}$ formalism (Schmidt 1968, Felten 1976) and the maximum 
likelihood method STY (Sandage, Tammann \& Yahil 1979), in order to 
represent our data with a Schechter (1976) parameterization. 
In the $1/V_{max}$ analysis, for any given redshift bin  ($z_1,
z_2$) a maximum volume is assigned to each object. This volume is calculated 
between $z_1$ and $z_{up}$, the latter being the minimum between 
$z_{2}$ and $z_{max}$,
i.e. the maximum redshift at which this galaxy would have satisfied the
magnitude limit ($K_s<20$) of the survey. 
We use the model tracks in order to compute $z_{max}$. 
Since the STY method determines the 
shape of the LF but not its overall normalization, we have 
normalized the STY LF by matching the galaxy number counts observed in each 
redshift bin. We have checked and tested the reliability of our results with the
independent softwares developed by Zucca et al. (1997) and by Poli et al.
(2001), always finding a
good agreement between the different methods.

Because of the relatively small number of objects in each redshift bin
and only a part of the luminosity function can be recovered, the Schechter 
parameters ($\alpha$, $M^*$ and $\phi^*$) derived from our STY analysis are 
not very well constrained by our data.
In particular, the uncertainty on the faint end slope $\alpha$
increases with redshift because of the increase with
redshift of the minimum observable luminosity. A sample
reaching $K_s \sim 21.0$--$21.5$ would be needed to better constrain
this parameter at $z > 1$. The results will be discussed in 
Section \ref{sec:LFz}.

\subsection{The treatment of the unidentified objects}\label{sec:unid}
 
The treatment of the unobserved or spectroscopically unidentified sources 
is not a major problem because of the high spectroscopic redshift 
completeness in our survey.
Moreover, as discussed in Sect. \ref{sec:sample}, tested and 
reliable photometric 
redshifts were derived for most of the faint unidentified or unobserved 
objects, leading to an almost negligible number of objects (9 in total) 
without any redshift information (see Paper III).

The small uncertainty introduced by the unobserved or unidentified objects 
has been addressed in two different ways. 

We first use only the spectroscopic sample and take into account the 
remaining redshift incompleteness (72 galaxies) by applying weights to 
each galaxy with spectroscopic redshift both in the 
$V_{max}$ analysis (Avni \& Bahcall 1980) and in the standard STY 
formulation (Heyl et al. 1997, cf. also Zucca, Pozzetti \& Zamorani 1994). 
The weighting corrections are computed on the basis of the
fraction of galaxies with spectroscopic redshift in different
regions of the $(R - K_s) - K_s$ plane. This weighting scheme
{\it assumes} that unidentified objects have the
same redshift distribution as the spectroscopically identified
with similar $K_s$ magnitudes and $R-K_s$ colors. While this scheme
can be reasonably applied over most of the $(R - K_s) - K_s$
plane, the weights are highly uncertain for the optically
faintest and reddest galaxies, most of which do not have a
spectroscopic redshift. Since the photometric redshifts of
these objects are statistically higher than those sampled
by the measured spectra, this method would not allow to 
correctly {\it estimate} the luminosity function
in the highest redshift bin.

\begin{figure}
\resizebox{\hsize}{!}{\includegraphics{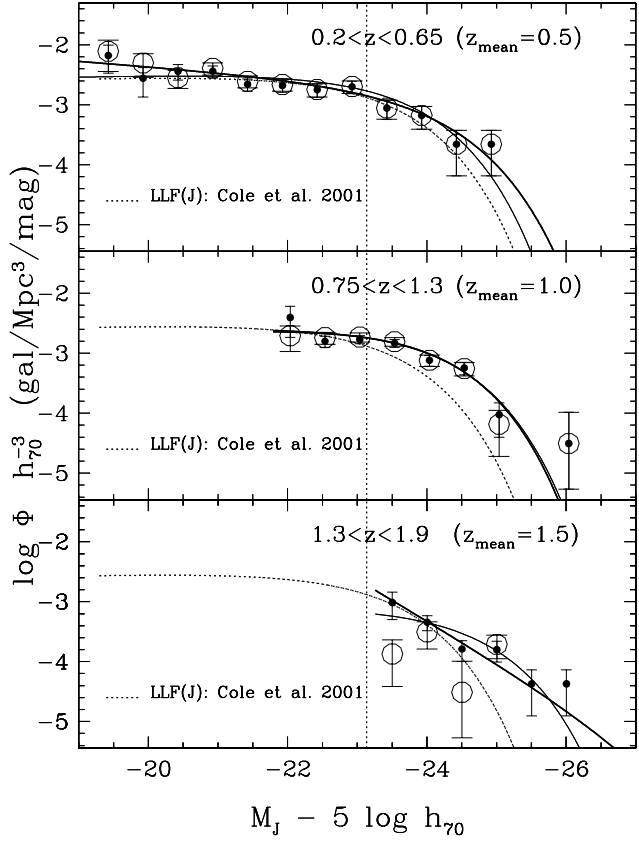}}
\caption{\footnotesize
The rest-frame $J$-band Luminosity Function in the redshift bins: 
$z_{mean}\simeq0.5$ (top panel) $z_{mean}\simeq1.0$ (middle panel) and
$z_{mean}\simeq1.5$ (bottom panel).
Points have been derived from $1/V_{max}$ analysis (open circles
using only spectroscopic $z$ and weighted incompleteness corrections, 
while filled small dots using both spectroscopic 
and photometric $z$, see Section \ref{sec:unid}),
while solid curves are the LF Schechter fits derived from the STY maximum
likelihood analysis (thin solid lines are the fits obtained fixing the
$\alpha$ parameter at the local value, see text).
The dotted curves and vertical dotted lines show the local LF in the $J$-band
and $M^*_J$ at $z=0$ from Cole et al. (2001).
}
\label{fig:LFJz}
\end{figure}

\begin{figure}
\resizebox{\hsize}{!}{\includegraphics{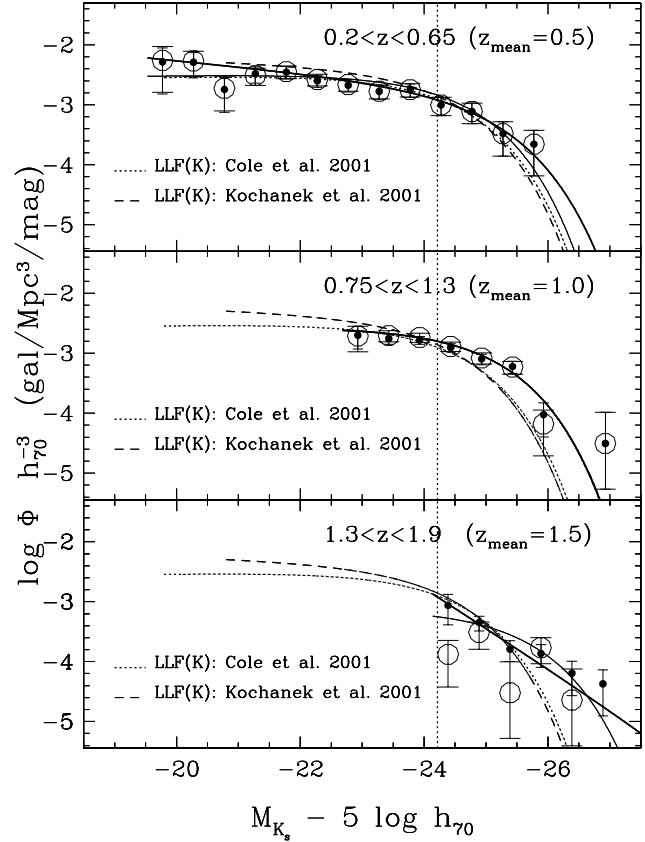}}
\caption{\footnotesize
The same as in Fig. \ref{fig:LFJz}, but for 
the rest-frame $K_s$-band.
The dotted curves and vertical dotted lines show the local LF in the $K_s$-band
and $M^*_K$ at $z=0$ from Cole et al. (2001), while dashed
curves show the local LF from Kochanek et al. (2001).
}
\label{fig:LFKz}
\end{figure}

As a second approach, we therefore base the ``best" estimate of the
luminosity function on the sample of galaxies with spectroscopic plus 
photometric redshifts. In this case, the 9 objects 
without any redshift information (because of the absence of the extended
multi-band photometry) 
have an almost negligible effect. To account for them we correct the STY 
normalization (from 1\% to 8\% in the 3 redshift bins) 
according to the fraction of such objects in each bin.
The results from the two methods (Fig. \ref{fig:LFJz} and \ref{fig:LFKz})
are completely consistent with each other
in the first two redshift bins, while
some discrepancies are present in the highest redshift bin 
because, as mentioned above, the weighting correction function is unable
to properly take into account the optically faintest and reddest galaxies. 
Furthermore we tested that the uncertainties 
in the photometric redshifts did not affect significantly our results.
We extracted random samples from the original photometric redshift 
catalog using the measured dispersion. 
The derived fluctuations in
each magnitude bin of the LFs resulted smaller than Poisson errors 
and in general the simulated LFs are 
consistent with the ``best" estimate also in the higher redshift bin 
with no significant systematic effects.
In the following sections we will discuss the ``best" LFs derived using 
spectroscopic plus photometric redshifts.

\section{The cosmic evolution of the near-IR Luminosity Function}
\label{sec:LFz}

Thanks to the high statistical significance and completeness of our sample 
it is possible to investigate the evolution of the near-IR LF over
a wide range of cosmic time.
Figs. \ref{fig:LFJz} and \ref{fig:LFKz} show the $1/V_{max}$ and 
the maximum likelihood (STY) results
for the LFs in the selected redshift bins 
and in the $J$- and $K_s$-bands, respectively.

We find that the luminosity functions are
fairly well fitted by Schechter functions in the first two redshift bins.
The magnitude ranges and the best-fit Schechter parameters are summarized in 
Table \ref{tab:LFP}, with the uncertainties derived from the projection of 
the 68\% confidence ellipse.
In the highest redshift bin, due to the bright and limited range in magnitude,
our data are well fitted with
a single power law ($\alpha\simeq-2.7^{+0.5}_{-0.7}$).
This does not mean that a Schechter function can not be a good representation 
of the luminosity function also in this redshift bin.
This is shown by the thin solid lines in the lower panels of Figs.
\ref{fig:LFJz} and \ref{fig:LFKz}, which represent fits with a Schecther 
function obtained by fixing the faint end slope $\alpha$ at the local
value. However, because of the bright limit in absolute magnitude
at these redshifts, which reduces significantly the range of sampled
luminosities, the Schechter parameters are very poorly determined
by these data.  For this reason the Schechter parameters for
this redshift bin are not reported in Table \ref{tab:LFP}.
We point out that a direct comparison of the Schechter parameters 
could be quite misleading
because the parameters are correlated. 

\begin{table}[t]
\begin{flushleft}
\caption{Galaxy luminosity evolution at different redshift }
\protect\label{tab:dMz}
\begin{tabular}{cccc}
\noalign{\smallskip}
\hline
\noalign{\smallskip}
\multicolumn{1}{c}{Band} & 
\multicolumn{1}{c}{$\Delta M(z=0.5)$} & 
\multicolumn{1}{c}{$\Delta M(z=1.0)$} &
\multicolumn{1}{c}{$\Delta M(z=1.5)$} \\
\noalign{\smallskip}
\noalign{\smallskip}
 $J$     & $-0.23^{+0.16}_{-0.17}$ & $-0.69^{+0.12}_{-0.12}$ & $-1.17^{+0.22}_{-0.24}$ \\ 
\noalign{\medskip}
 $K_s$   & $-0.11^{+0.17}_{-0.18}$ & $-0.54^{+0.12}_{-0.13}$ & $-1.07^{+0.23}_{-0.27}$ \\ 
\noalign{\smallskip}
\hline
\end{tabular}
\end{flushleft}
\end{table}

In order to trace the evolution down to $z\sim0$, we compared the observed
LFs as derived from our sample to the Local Luminosity Function (LLF)
by Cole et al. (2001) in the $J$ and $K_s$ bands (Figs. \ref{fig:LFJz}, 
\ref{fig:LFKz}) and by Kochanek et al. (2001) in the $K_s$-band. 
We then explored two possible evolutionary scenarios: luminosity or density 
evolution.

Our analysis (both $V_{max}$ and STY) shows that the 
data (Figs. \ref{fig:LFJz} and \ref{fig:LFKz})
are consistent with a mild evolution from $z=0$ to $z\simeq1.5$ both in 
$J$- and $K_s$-bands. In particular 
the faint-end of the LFs ($L<L^*$) is consistent with the
local estimates, with no statistically significant evidence for a change 
either in the slope or
normalization up to $z<1.3$ (consistently with Cohen 2002).
At higher redshift this part of the luminosity function is not
well sampled by our data,
but also in this bin a Schechter function with
local faint-end slope is consistent with the
data (Figs. \ref{fig:LFJz} and \ref{fig:LFKz}).
Viceversa, it is interesting to note that the 
density of luminous galaxies (e.g. $M_{K_s}-5$ log$ h_{70}<-25.5$)
is significantly higher than locally.

Because of the statistical correlation between the Schechter parameters
($\alpha$, $M^*$ and $\Phi^*$), in order to estimate  the
luminosity and/or density evolution within our sample and in comparison
with the local LFs, we have 
fixed the $\alpha$ parameters of our LFs to the same values observed
locally (Cole et al. 2001),
$\alpha_J=-0.93$ and $\alpha_{K_s}=-0.96$
(thin solid lines in Figs.  \ref{fig:LFJz}, \ref{fig:LFKz}).
We are allowed to do this because all our best fit $\alpha$ values are
consistent with the local ones (see Table \ref{tab:LFP}).
Assuming local $M^*_J=-23.13+5$log$h_{70}$ and 
$M^*_{K_s}=-24.21+5$log$h_{70}$ from Cole et al. (2001) 
we find a {\it luminosity evolution} at $z\simeq1$ of the order 
of $\Delta M_J \simeq -0.69\pm0.12$  and $\Delta M_K \simeq -0.54\pm0.12$
with the normalizations 
($\phi^*$) in the first two redshift bins consistent with the local values, 
within from 3 to 20\%. 
At $z>1.3$ there is an indication of an even higher luminosity evolution 
($\Delta M_J \simeq -1.17$, $\Delta M_K \simeq -1.07$), 
while the $\phi^*$ values 
decrease by a factor $4$--$5$ with respect to the local value. This last
result should be taken with caution because of the lower statistics in the
highest redshift bin and the smaller range in luminosity covered.

\begin{figure}
\resizebox{\hsize}{!}{\includegraphics{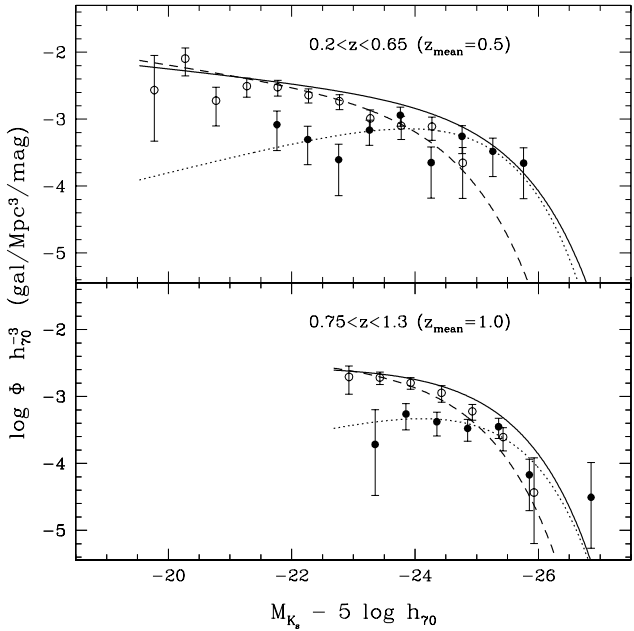}}
\caption{\footnotesize
Rest-frame $K_s$-band Luminosity Function 
in two redshift bins: $z_{mean}\simeq0.5$ (top panel)
$z_{mean}\simeq1.0$ (bottom panel).
Points and lines derive from $1/V_{max}$ and STY analysis for early 
(filled symbols and dotted lines) and late (open symbols and dashed lines) 
galaxies respectively. The solid curves are the STY LF for the total sample 
in the $K_s$-band as in Fig. \ref{fig:LFKz}.
}
\label{fig:LFKz_el}
\end{figure}

The luminosity evolution estimated from STY analysis are summarized
in Table \ref{tab:dMz} with the formal statistical $1 \sigma$ confidence limits.
The uncertainties introduced by the k-correction, discussed in Section 
\ref{sec:LIR}, are always less than the statistical errors, and less than 0.03 
and 0.10 magnitudes at $z<1.3$ and $z\simeq1.5$, respectively. 

We recall here that Cimatti et al. (2002b) discussed the photometric selection 
effects present in the K20 sample and showed that, on average, the total flux
of spirals and ellipticals with $L\leq L^*$ are underestimated by 0.1
and 0.25 magnitudes respectively, while at higher $L$ the flux lost
for elliptical galaxies could be even higher. For this reason we are
confident that our estimate of a positive luminosity evolution is quite 
conservative. The above estimates of the luminosity evolution could increase 
on average
by about $-0.2$ magnitudes if we take into account this underestimate of
the total flux.
\section{The Luminosity Function by spectral or color type}
\label{sec:type}

In order to investigate the role of different galaxy populations,
we divide the K20 sample in two subsamples
on the basis of the spectroscopic classification, i.e. early
type galaxies without strong emission lines and late type galaxies with 
emission lines,
and  study the near-IR LF for the different populations in the two
redshift bins at $z_{mean}\simeq0.5$ and $1$. The high fraction of photometric 
redshifts without any spectroscopic information does not allow us to extend 
such studies to the highest redshift bin.
We take into account the spectral incompleteness using two weighting 
functions (see Sect. \ref{sec:unid}) derived for each spectral type and 
correcting the overall normalization in the STY method using the color 
information of unclassified objects. 

The results are shown in Fig. \ref{fig:LFKz_el}. 
Our analysis indicates that the faint-end slope of the LFs at these
redshifts shows the same general dependence on galaxy
spectral type as that found in numerous studies of the local LF. Specifically,
the faint-end slope is much steeper for late/emission line galaxies
than for early type galaxies. A similar result has been found by Cohen
(2002) in the HDF--north.
More important, we find, for the first time, that at $z_{mean}\simeq0.5$,
early type galaxies clearly dominate the bright-end of the Luminosity
Function, in agreement with local observations (Kochanek et al. 2001).
A similar result is also visible at $z_{mean}\simeq1.0$. 
Some of the previous surveys, selected in the optical bands with
follow-up $K$-band observations (e.g. Cohen 2002),
present a bias against identifying red early type galaxies
at high redshift because of substantial 
incompleteness of EROs in the samples.

The overall density 
of early type galaxies at $z\simeq0.5$ is consistent with the local estimate
(Kochanek et al. 2001), while at $z\simeq1$,
their number density shows at most a small decrease ($<30\%$), 
consistent with recent results from Im et al. (2002). 
This decrement could be due both to the existence of a population 
of galaxies morphologically classified as
early-type but with blue colors due to some recent episodes of star formation 
(Menanteau, Abraham \& Ellis 2001, Im et al. 2001) 
and to our spectral incompleteness,
more severe for red/early type galaxies.

Similar results are obtained if we divide the galaxies in two samples 
according to their $R-K_s$ colors (we adopt the color of Sa galaxies to
divide the sample in red and blue). 

We conclude that {\it red and early type galaxies dominate the 
bright-end of the Luminosity Function already at early epoch 
and that their number density shows at most a small decrease ($<30\%$) up
to $z\simeq 1$}. 

\begin{figure}
\resizebox{\hsize}{!}{\includegraphics{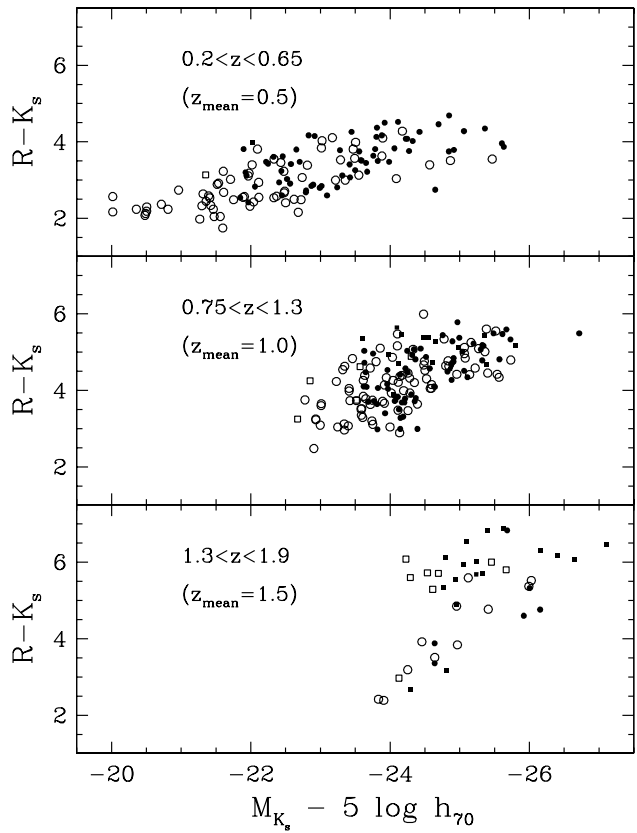}}
\caption{\footnotesize
$R-K_s$ colors vs. rest-frame absolute $K_s$ magnitudes 
for  K20 galaxy sample in 3 different redshift bins.
Circles represent galaxies identified spectroscopically, while 
squares are unidentified or unobserved galaxies plotted at $z=z_{phot}$.
In each redshift bin
empty and filled symbols refer to galaxies with $z<z_{mean}$ and 
$z\ge z_{mean}$,
respectively.
}
\label{fig:RK_MK}
\end{figure}

Fig. \ref{fig:RK_MK} shows the optical/near-IR colors versus near-IR
luminosities in the three redshift bins.
While the magnitude limit of the survey,
$K_s<20$, corresponds to different low luminosity limits in the different
redshift bins,  
there appears to be a correlation of the optical/near-IR colors 
with near-IR luminosities.
In first approximation this correlation can be interpreted as a
correlation of
age and/or specific star formation rate with stellar mass, with 
the most ``massive"
galaxies being ``old'' and the ``low-mass" galaxies being instead dominated
by young stellar populations. This is now well established for local
galaxies (Gavazzi et al. 1996, Boselli et al. 2001, Kauffmann et al.
2002) and we extend it to $z\sim1.5$. 
Fig. \ref{fig:RK_MK} (see empty and filled circles representing galaxies 
with $z<z_{mean}$ and
$z\ge z_{mean}$) shows that the observed correlation is not induced by a
redshift effect within the bins.
In Sect. \ref{sec:models} we compare the observed correlation with
HMM predictions.

\begin{table}[t]
\begin{flushleft}
\caption{Luminosity density$^a$
at different redshift }
\protect\label{tab:LdVz}
\begin{tabular}{ccccc}
\hline
\noalign{\smallskip}
\multicolumn{1}{c}{redshift} & 
\multicolumn{1}{c}{log ${\rho_J}^b$} & 
\multicolumn{1}{c}{log ${\rho_J}^c$} & 
\multicolumn{1}{c}{log ${\rho_{K_s}}^b$} & 
\multicolumn{1}{c}{log ${\rho_{K_s}}^c$} \\
\noalign{\smallskip}
\noalign{\smallskip}
 $0.5$ & $20.20$ & $20.20^{+0.01}_{-0.02}$ & $20.19$ & $20.17^{+0.01}_{-0.02}$\\
\noalign{\medskip}
 $1.0$ & $20.25$ & $20.28^{+0.05}_{-0.03}$ & $20.21$ & $20.26^{+0.05}_{-0.03}$\\
\noalign{\medskip}
 $1.5$ & $>19.90$ & $-$ & $>19.88$ & $-$ \\ 
\noalign{\smallskip}
\hline
\noalign{\smallskip}
\end{tabular}

$a$: in $h_{70}$ W/Hz/Mpc$^3$ units

$b$: observed

$c$: LF-corrected (see text).
\end{flushleft}
\end{table}

\begin{figure}
\resizebox{\hsize}{!}{\includegraphics{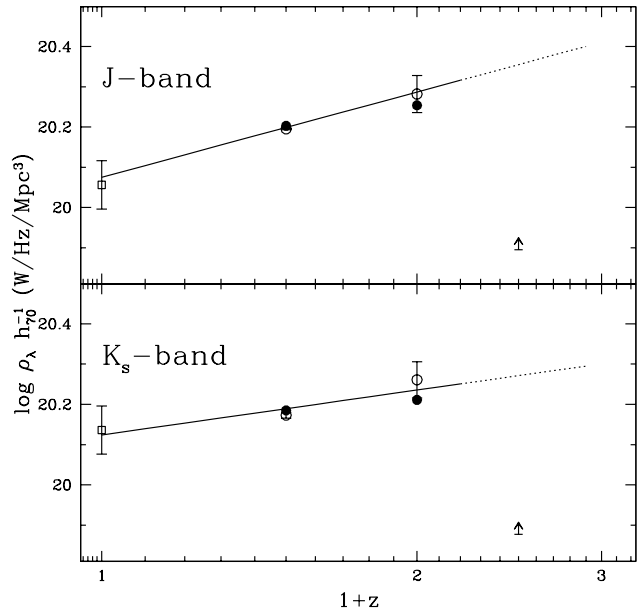}}
\caption{\footnotesize
Comoving luminosity density in $J-$ (top panel) and $K_s$-bands (bottom panel).
The filled circles are the value derived directly from observations (see
text), while open circles are the ``LF-corrected'' estimates. The solid lines
show the best-fit power laws. The open squares are from Cole et al. (2001).
}
\label{fig:LdVz}
\end{figure}

\section{The near-IR luminosity density evolution}
\label{sec:LdVz}

Tracing the cosmic emission history of the galaxies at different wavelengths
offers the prospect of an empirical determination of the global
evolution of the galaxy population. Indeed, it is independent of the 
details of galaxy evolution and depends mainly on the star formation 
history of the universe (Lilly et al. 1996, Madau, Pozzetti \& Dickinson 1998).
Various attempts to reconstruct the cosmic evolution of the comoving luminosity 
density have been made previously mainly in the UV and optical bands (Lilly 
et al. 1996, Cowie et al. 1999). Our survey offers 
the possibility to investigate it in the near-IR using a LF extended over
a wide range in luminosity.

We estimate the comoving luminosity density directly from observed 
sources using the $V_{max}$ formalism (see for details Lilly et al. 1996).
The major source of uncertainty is the contribution from galaxies fainter
than the magnitude limit $K_s<20$. We have therefore estimated also
the ``LF-corrected" luminosity density using the best-fit LF from STY
method and extending it to fainter luminosity 
(we adopt $0.02 L^*(z)$ as our lower limit in luminosity for this
extrapolation).
A formal uncertainty in this procedure was estimated by considering
the range of acceptable Schechter parameters values (see Table
\ref{tab:LFP}).

Fig. \ref{fig:LdVz} shows luminosity densities derived in $J$- and
$K_s$-band as a function of redshift. Up to $z<1.3$ the uncertainties
due to faint galaxies and LF parameters are small. At higher redshift the
uncertainties in the LF parameters do not allow to constrain the luminosity 
density values. 
We could therefore only estimate a lower limit to the luminosity density 
using the observed value.

We find a slow evolution with redshift of the observed  near-IR luminosity
density, consistent with the results by Cohen (2002) in the HDF-north,
even if she had to apply a large correction to her data
because of the disappearance of absorption galaxies at high redshift in
her sample as they became EROs.
Using the observed local luminosity densities derived from Cole et al. (2001), 
the luminosity density evolution up to $z\le1.3$ is well represented by 
a power law, 
$\rho_\lambda(z)= \rho_\lambda(z=0) (1+z)^{\beta(\lambda)}$. We find 
$\beta(J)\simeq0.70$ and $\beta(K_s)\simeq0.37$ (Fig. \ref{fig:LdVz}). 

The near-IR luminosity density evolution is much slower
than that found in the UV and optical bands 
($\beta=3.9$--$2.7$ from 0.28 to 0.44 $\mu$m by Lilly et al. 1996 and 
 $\beta=1.5$ at 0.15, 0.28 $\mu$m by Cowie et al. 1999, for $\Omega_m=1$).
Indeed, while the optical luminosity density evolution is mainly related to the 
star formation history, the evolution of the near-IR luminosity density is
more closely related to the stellar-mass density. 
The local stellar mass density, derived by Cole et al. (2001) is
$\Omega_{stars} = (3.7 \pm 0.6) \times 10^{-3} h_{70}^{-1}$.
If we adopt the stellar mass-to-light ratio 
for the galaxy spectral models which best match colors and
spectral types for each galaxy,
the mean ${\cal M}_{stars}/L_K$ in our sample becomes $0.63$ and $0.54$  
at $z=0.5$ and $1$, with small variations due to the model parameters
adopted (see Sect. \ref{sec:LIR}).
With these values of ${\cal M}_{stars}/L_K$ we derive  
$\Omega_{stars}=(2.0\pm 0.1) \times 10^{-3} h_{70}^{-1}$ and
$\Omega_{stars}=(2.1\pm 0.3) \times 10^{-3} h_{70}^{-1}$ at $z=0.5$ and
$1$ respectively.
This analysis suggests that the evolution of the stellar mass density
is relatively slow with redshift, with a decrease of about
a factor $1.8\pm0.4$ from $z=0$ to  $z\simeq1$, 
consistently with recent results from Dickinson et al. (2002a). 

At $z>1$ we can only give a lower limit to the 
near-IR luminosity density. As discussed by Madau, Pozzetti, Dickinson (1998) 
(cf. also Pozzetti \& Madau 2001), a robust determination of 
the near-IR luminosity density at
$z>1$ could be fundamental to disentangle between different cosmic 
histories of star formation (SFH). In fact, SFHs peaked at intermediate 
redshift ($1<z<2$) predict a decrease in the near-IR luminosity 
density at $z>1$--$2$, 
while if 50\% of the stars formed at $z>2$ (similar to the PLE adopted here), 
the corresponding near-IR luminosity density flattens at $z>1$ 
without a strong decrease (cf. Fig. 6 in Pozzetti \& Madau 2001).
A detailed comparison of luminosity and stellar-mass density with model
predictions and empirical star formation histories will be further
investigated in a forthcoming paper.

\begin{figure*}[ht] 
\resizebox{0.5\hsize}{!}{\includegraphics{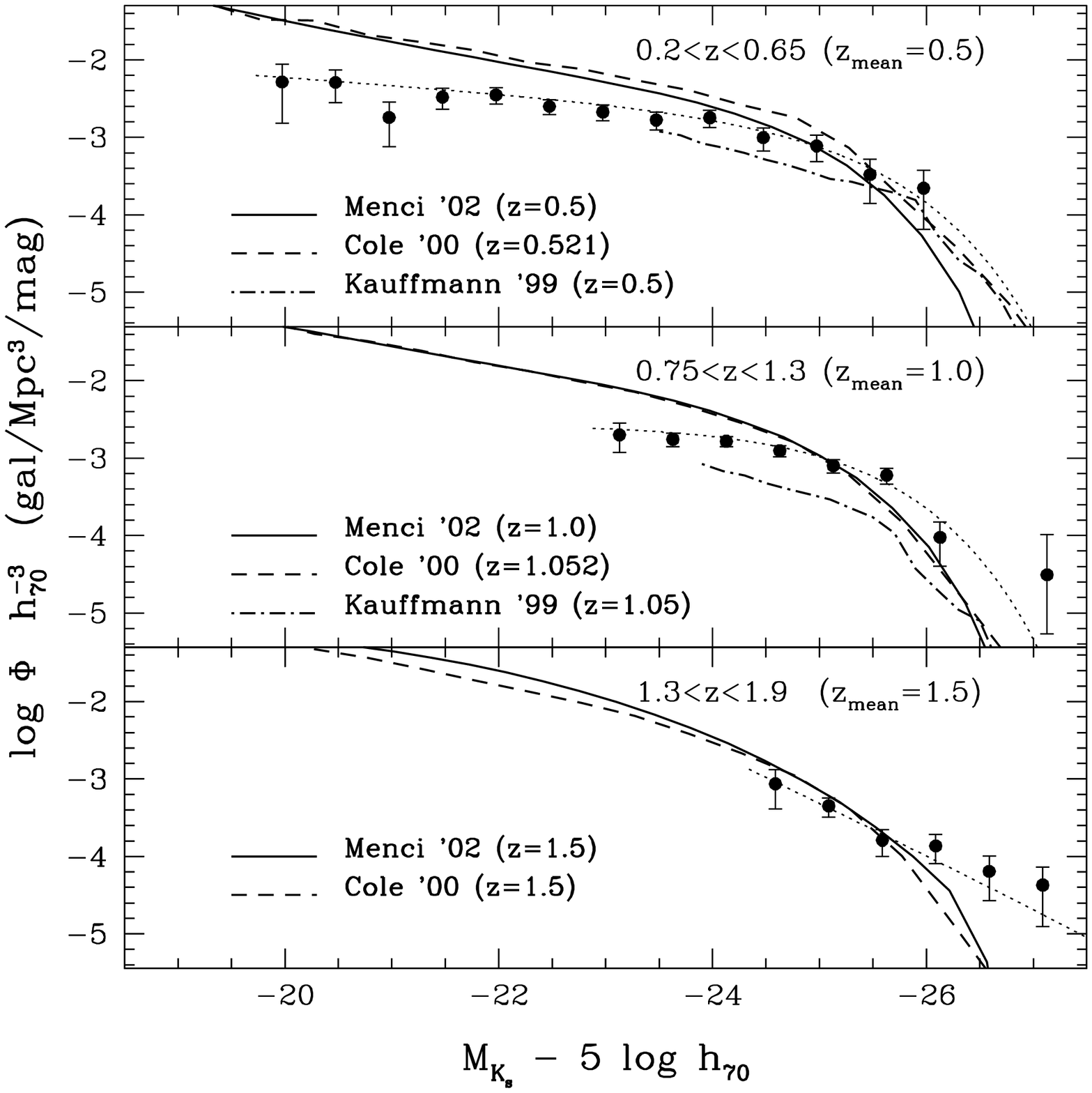}}
\resizebox{0.5\hsize}{!}{\includegraphics{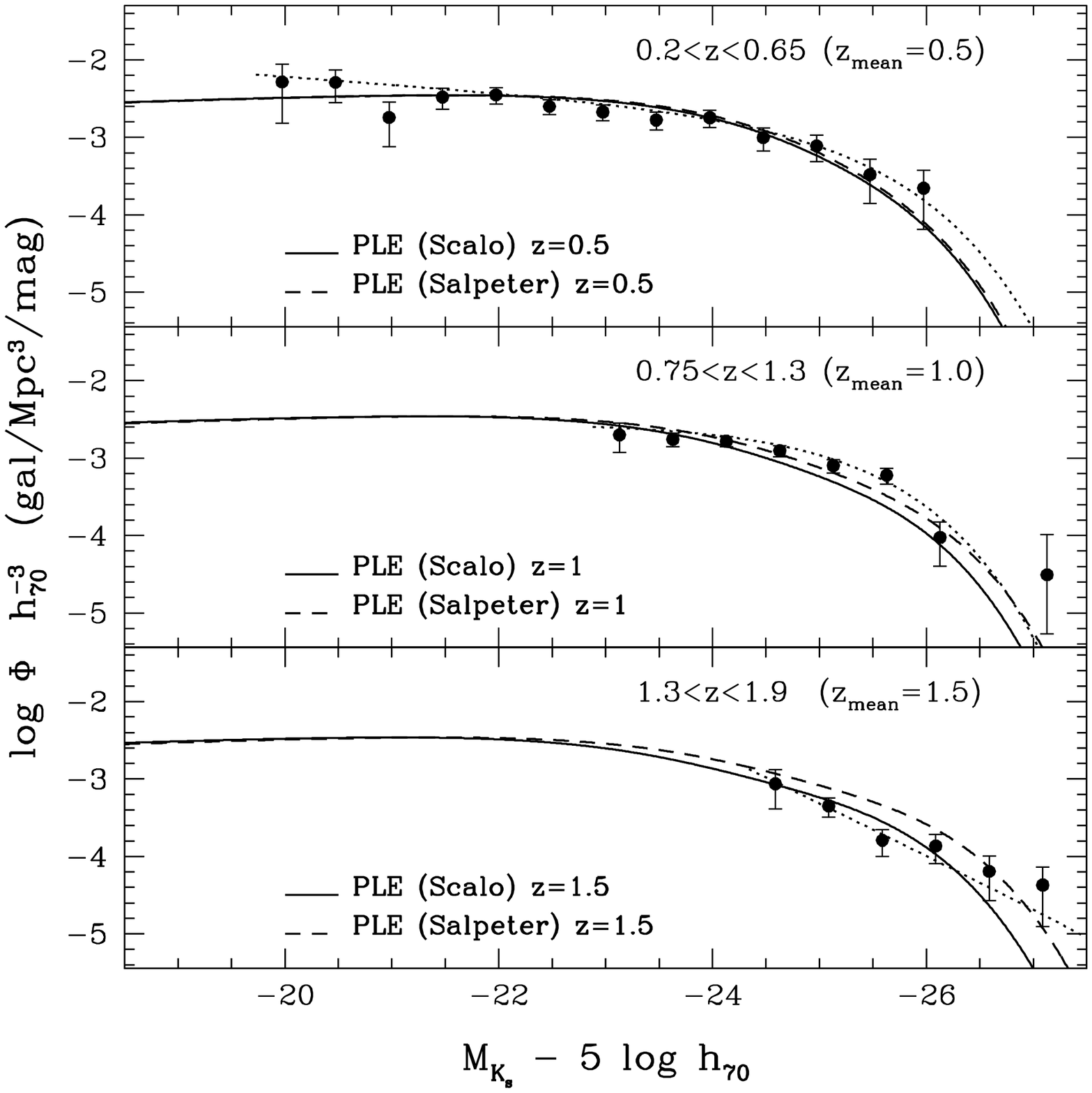}}
\caption{\footnotesize  
The rest-frame $K_s$-band Luminosity Function in the redshift bins: 
$z_{mean}\simeq0.5$ (top panels) $z_{mean}\simeq1.0$ (middle panels) and
$z_{mean}\simeq1.5$ (bottom panels) compared to PLE (right panels)
and HMM (left panels) models. We have used the models closer to $z_{mean}$
or interpolated between different redshifts.
Data points derive from $1/V_{max}$ analysis, while dotted curves are 
the LF Schechter fit derived from STY analysis. 
We have corrected the data on average by $-0.2$ magnitude because photometric
selection effects (see text).
}
\label{fig:LFKM}
\end{figure*}

\section{The comparison with model predictions}
\label{sec:models}

We have compared our LF with the predictions from two competitive 
scenarios of galaxy evolution: the Hierarchical Merging Models (HMMs) 
and the Pure Luminosity Evolution (PLE) (Fig. \ref{fig:LFKM}). 
For the HMMs, we used the 
predictions by Cole et al. (2000, C00, kindly provided us by C. 
Baugh), Menci et al. (2002, M02), and GIF (Kauffmann et al. 1999).
For the GIF simulations, which combine the large high-resolution N-body
simulations with semi-analytical models, 
we derive the LF at a magnitude above which it can be considered
complete in luminosity given the 
mass limit of the simulated galaxy sample 
($2\times 10^{10}$ $h_{70}^{-1} M_\odot$).
For the PLE parameterization, we adopted the predictions of Pozzetti 
et al. (1996, 1998, PPLE) (for details see also Paper IV).
 
For a more reliable comparison with models, we corrected the observed 
magnitudes by $-0.2$ mag, in order 
to take into account the average loss of total flux 
in the K20 sample (see Papers III-IV).

As discussed in Paper IV at low-$z$ ($z=0.5$) the HMMs by Cole et
al (2000) and by Menci et al. (2002) predict many more galaxies than
observed (cf. Fig. 3 in Paper IV). 
From the present study (upper panel of Fig. \ref{fig:LFKM})
we can therefore conclude that such excess is due to low luminosity, 
``low-mass`` objects, which dominate the steep faint-end in the HMM LF.
At $z=1$, the HMMs predict even steeper LFs because low-mass galaxies
become more numerous in the hierarchical scenario at higher redshift. 
This is in contrast with the LF derived from our data at $z=1$, which 
does not show a similar behaviour. 
These discrepancies are consistent with 
previous results both in the local universe (cf. Fig. 1 in 
Baugh et al. 2002) and at high redshift ($z<1$ and $z\sim 3$) for the B and UV
rest-frame LF, respectively (Poli et al. 2001, Somerville et al. 2001). 

\begin{figure}
\resizebox{\hsize}{!}{\includegraphics{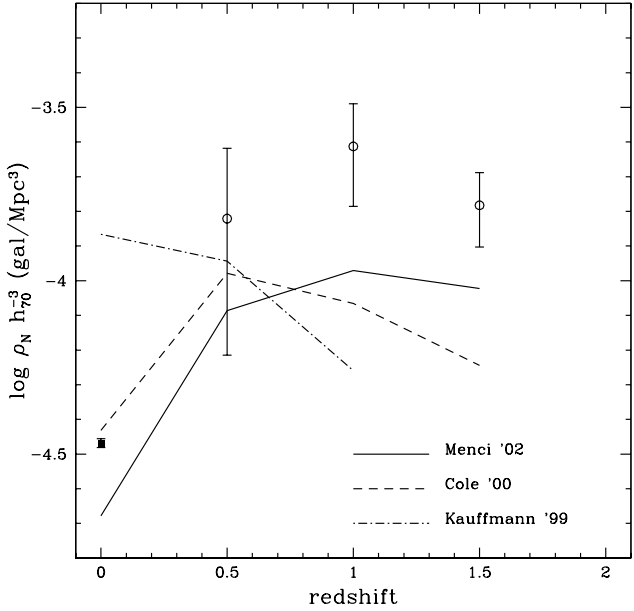}}
\caption{\footnotesize
Comoving number density of luminous galaxies ($M_{K_s}<-25.5+5$
log$h_{70}$) in the $K_s$-bands.
The open circles are the values derived directly from observations,
while the filled square is from Cole et al. (2001). The different lines
show models as in figure \ref{fig:LFKM}.
}
\label{fig:NdVz}
\end{figure}

Also at the bright end of the LF the HMMs appear to be in disagreement 
with our data at $z\simeq1$, where they underpredict the
density of bright ($M_{K_s}<-25.5+5$log$ h_{70}$) galaxies with respect to 
our data. For example, our data at $M_{K_s}=-25.6+5 $log$ h_{70}$
and $z\simeq1$ are about a factor 2.6 higher
than the HMM predictions
and even higher at brighter magnitudes
if we consider the STY maximum likelihood fit.
In the highest redshift bin the excess over the predictions 
becomes significant at $M_{K_s}<-26 +5$ log$ h_{70}$.
In comparison, the GIF simulated catalogue at $z=1$ underpredicts the K20 LFs 
at all magnitudes.
As discussed by Kauffmann et al. (1999), the GIF model produces a factor
$2$--$3$ too few galaxies at magnitudes around $L^*$ 
already at $z=0$,  while it produces an excess of bright local galaxies.
This problem with respect to the local LF affects also the Cole et al. (2000) 
model (see Fig. 4 in Baugh et al. 2002).
Therefore, as discussed in Sect. \ref{sec:LFz}, while the 
density of luminous objects ($M_{K_s}-5$ log$ h_{70}<-25.5$)
is quite constant or mildly increasing with $z$ within our sample,
and higher than the local density, it 
is rapidly decreasing in the Cole et al. (2000) and GIF models at $z>0.5$ 
(Fig. \ref{fig:NdVz}) in clear conflict with our data. 
The problem of a negative evolution with redshift of luminous objects 
does not affect instead the Menci HMM
model, even if also this model significantly and systematically underpredicts
their number density at all redshifts.

\begin{figure}
\resizebox{\hsize}{!}{\includegraphics{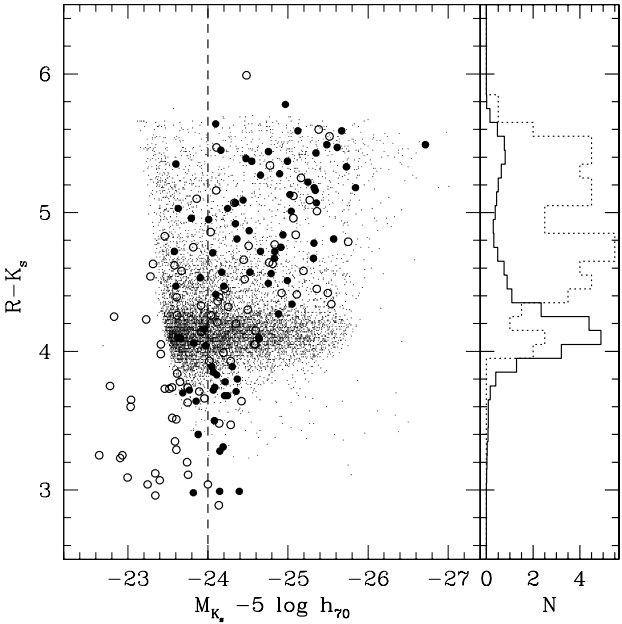}}
\caption{\footnotesize
{\it Left panel:} $R-K_s$ colors vs. rest-frame absolute $K_s$ magnitudes 
for $z=1.05$ GIF simulated catalogue (small dots)
and data (circles) at $0.75<z<1.3$ ($z_{mean}\simeq1$)
(empty and filled circles refer to $z<1$ and $z>1$ respectively).
Vertical dashed line represents approximately the completeness magnitude
limit of GIF catalogue corresponding to its mass limit (see text).
{\it Right panel:} Color distribution of high-luminosity
galaxies ($M_{K_s}-5$ log $h_{70}<-24.5$) observed (dotted line) and
simulated (continue line), normalized to the same comoving volume. 
}
\label{fig:RK_MK_H}
\end{figure}

In addition to the comparison of the data and predictions for the LF, it
is also important to verify whether the bright $L>L^*$ galaxies in the
K20 survey have colors consistent with the HMM predictions.
Fig. \ref{fig:RK_MK_H} shows the comparison between the $z=1.05$ GIF simulated
catalogue and data at $0.75<z<1.3$ ($z_{mean}\simeq1$). While the 
deficiency of simulated blue low-luminosity ($R-K_s<4.2$, $M_{K_s}-5$ log
$h_{70}>-23.5$) galaxies is due mainly to the mass limit of the GIF
catalogue,
a serious discrepancy emerges in the two distributions in the magnitude
range where the GIF catalogue is expected to be complete.
The GIF simulated catalogue shows an
excess of  blue high-luminosity galaxies ($R-K_s<4.4$, $M_{K_s}-5$ 
log $h_{70}<-24.5$) and a deficiency of red luminous galaxies
(see also the color distributions of high-luminosity
galaxies, $M_{K_s}-5$ log $h_{70}<-24.5$, in the right panel of Fig.
\ref{fig:RK_MK_H}, normalized 
to the same comoving volume).
This result is in agreement with the finding of a deficiency of EROs in 
the HMMs (Daddi et al. 2000, Firth et al. 2002, Smith et al. 2002),
but, since it is based not only on colors but also
on the luminosity of the galaxies, it points out in particular
a {\it clear deficiency of red luminous galaxies in the HMM predictions}.
This plot clearly shows that
a comparison of LF {\it and} colors is much more
powerful than that of the colors or LF alone and that 
the GIF $K_s$-band LF at $z=1$ is obtained
with a radically different mix of galaxies compared to the observed one.
While the bright end of the HMM LF at $z=1$ is dominated by actively
starforming galaxies, in the K20 sample it is dominated by the passively 
evolving galaxies. 
Given that blue galaxies suggest younger stellar population and lower 
${\cal M}_{stars}/L_K$ ratios, 
we expect that the failure of this model in reproducing our LF
(see Fig. \ref{fig:LFKM}) should be even stronger in term of density
of ``massive" galaxies at $z\sim1$ (cf. also the luminosity and stellar-mass 
functions in Baugh et al. 2002).

The right panels of Fig. \ref{fig:LFKM} show the comparison of our data
with the PLE predictions, computed with two different
IMF (Scalo and Salpeter). The overall agreement is
much better than with the HMM predictions even if some
discrepancies, but at much lower level of significance, are
present. In particular, in the highest redshift bin the
parameterization with the Salpeter IMF overpredicts the total
number of galaxies (see also Paper IV), while the Scalo IMF
(dominated by low/intermediate mass stars, which induce a lower
luminosity evolution compared to Salpeter at $z>1$)
fits well the density of galaxies at $M_{K_s} > -26.0 +5$ log$h_{70}$, 
but falls below the data for the brighter galaxies.
In Paper IV we found that in the context of 
PLE models also the observed redshift distribution of K20 galaxies is
better fitted assuming an IMF dominated by low/intermediate mass stars
(e.g. Scalo IMF) than a Salpeter IMF (see also Pozzetti et al. 1996 and
Broadhurst \& Bouwens 2000).
PLE models with a flatter IMF (i.e. Salpeter-like) can be made
consistent with the data (redshift distribution and LF) only if the
rapid evolution induced by massive stars is sufficiently suppressed by
dust attenuation (Totani et al. 2001, see Paper IV). Indeed, such a
dusty phase is predicted by many models for the formation of high
redshift spheroids and could be associated to high redshift
SCUBA sources.

\section{Summary and discussion}
\label{sec:summary}

The cosmic evolution of the $K_s$-band selected field galaxy population 
has been studied over the redshift interval $0.2<z<1.9$ using about 
500 galaxies from the K20 spectroscopic survey. The sample spans a 
wide range in redshift and look-back time and allows to study the evolution of 
the near-IR Luminosity Function both within the sample and in comparison to 
the local population.  

We take advantage of the near-IR selection (in particular in the
$K$-band), in which the k-corrections are relatively invariant 
to galaxy type and relatively small also at high redshift (Cowie et 
al. 1994), and the dust extinction effects are less severe than in 
optical samples. 

We derived the near-IR luminosity function in the rest-frame $J$ and 
$K_s$-band in three redshift bins ($z_{mean}\simeq0.5,1.0,1.5$).
The detailed analysis of the observed LF at different redshifts and the 
comparison with the local LF and with the predictions of galaxy formation
models provided the following results:

1- A mild evolution is observed both in the $J$ and $K_s$ Luminosity 
Functions to $z\simeq1.5$, in agreement with previous 
indications by Cowie et al. (1996), Cohen (2002) and Feulner et al.
(2003). 
There is no evidence of a steepening of the faint-end LFs up to $z\sim1.3$. 
In particular, the faint-end ($L<L^*$) is 
consistent with the local estimates up to $z<1.3$, while at the bright end the 
data show a luminosity evolution of about $\Delta M_J \simeq -0.69\pm0.12$ and 
$\Delta M_K \simeq -0.54\pm0.12$  at $z\simeq1$.
The density of luminous galaxies ($M_{K_s}-5$ log$ h_{70}<-25.5$) is 
relatively constant or mildly increasing with redshift within our sample
and higher than locally at all redshifts.

2- Pure density evolution cannot reproduce the observed LF at $z\le 1.3$. 

3- Red and early-type galaxies clearly dominate the bright-end of the Luminosity
Function at $z\simeq0.5$ and a similar trend is also visible at $z\simeq1$, 
showing that such systems were already
in place and fully assembled at that cosmic epoch.
Their number density shows at most a small ($<30\%$) decrease up
to $z\simeq 1$. 

4- The evolution of the rest-frame near-IR comoving luminosity densities up to 
$z\simeq1$ can be described by power laws, $\rho_\lambda(z)= \rho_\lambda(z=0) 
(1+z)^{\beta(\lambda)}$, with $\beta(J)\simeq0.70$ and $\beta(K_s)\simeq0.37$. 
Such an evolution is much slower than those observed in the UV and optical
bands.

5- The hierarchical merging models overpredict the LF at low luminosity 
at all redshifts, whereas {\it they underpredict the density of 
high-luminosity galaxies at $z>1$}.
HMMs by Kauffman et al. (1999) and Cole et al. (2000) overpredict high
luminosity galaxies at $z=0$, and predict a negative density
evolution of the bright end of the LF at $z>0.5$ which is not observed.

6- The PLE predictions are in rather good agreement with the mild luminosity 
evolution observed up to $z\simeq1.5$. 

7- There appears to be a clear
correlation of the optical/near-IR colors, i.e. in first approximation
the specific star formation rate and ages, with near-IR luminosities, 
i.e. the stellar mass. This correlation suggests that the most ``massive"
galaxies are ``old'', while the ``low-mass" galaxies are instead dominated
by young stellar populations. 
The GIF model shows instead
a {\it clear deficiency of red luminous galaxies at $z\sim1$} compared to the
observations.

From the analysis of the LF and of the comoving near-IR luminosity
density, we can derive some indications on the evolution 
of the Galaxy Stellar-Mass Function (GSMF). 
A detailed analysis will be presented by Fontana et al. (2003). 
The present analysis suggests that the evolution of the GSMF and of the 
stellar mass density is slow with redshift up to $z\simeq 1.5$,
in contrast with the rapid evolution of the Galaxy Stellar Mass Function
expected in the hierarchical models at $z<2$ (cf. Baugh et al. 2002).
In our sample, using galaxy spectral models which match colors and
spectral types, 
we found that $\Omega_{stars}$ decreases by a factor about $1.8\pm0.4$ 
from $z=0$ to $z\simeq1$, mainly due to the decrease in the
mass-to-light ratio, which is however slower than in HMMs models. In at least
one of these models (GIF) the small adopted M/L ratio corresponds to a
color distribution of galaxies at $z\sim1$ clearly inconsistent with
data (see Fig. \ref{fig:RK_MK_H}).

Moreover, the fact that the bright-end of the Luminosity Function is dominated 
by red/early type galaxies and that the mean spectra of early type EROs 
at $z\simeq1$ are consistent
with an old stellar population (see Paper I), suggest that old and 
massive elliptical galaxies 
were already in place at $z\simeq1$ (see also Paper I) and,
{\it therefore, they should have formed their stars and assembled their mass
at higher redshift.} These results 
are in contrast with the current renditions of hierarchical models in 
which the bulk of massive elliptical galaxies forms through merging at low 
redshift ($z<1$--$2$)
and suggest that most of the merging which form elliptical galaxies in 
the hierarchical models should occur at higher redshift (say, $z>2$--$3$), 
followed by a pure ``luminosity-like" evolution. 
On the other hand, HMMs overpredict 
the density of low-luminosity, low mass, galaxies at $z<1$, as previously 
noted in the local universe (cf. Fig. 1 in Baugh et al. 2002) and in the B 
and UV rest frame bands at $z\ge1$ (Poli et al. 2001, Somerville et al. 2001,
Poli et al. 2003).

\begin{figure}[ht]
\resizebox{\hsize}{!}{\includegraphics{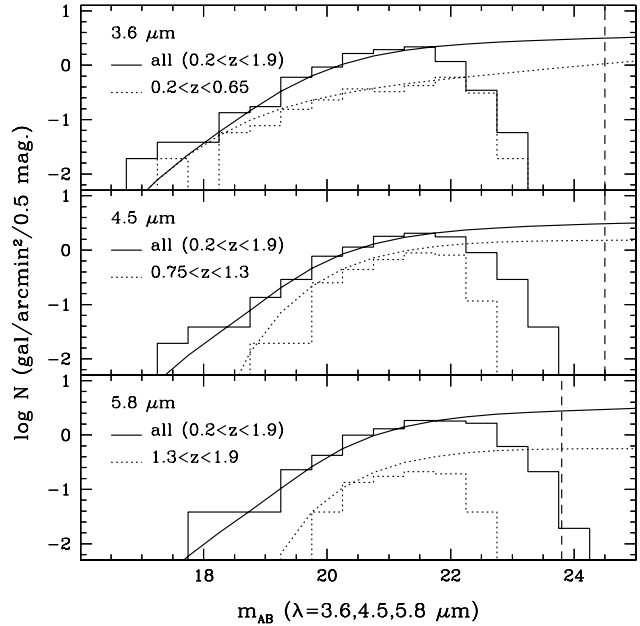}}
\caption{\footnotesize  
Predicted SIRTF+IRAC number densities for galaxies with $z<2$ 
derived from K20 sample (see text) as a
function of AB magnitudes at 3 different wavelengths 
($\lambda=3.6,4.5,5.8 \mu$m).
Histograms refer to $K_s<20$ galaxies in our sample in the indicated redshift 
ranges, while lines 
have been derived from K20 LFs extended to fainter luminosities (using
the flat slope in the highest $z$ range). 
Vertical dashed lines refer to GOODS flux limits at
the different wavelengths.
}
\label{fig:MIRz}
\end{figure}

Our results show the importance of sampling the faint-end of the LF in 
even deeper near-IR selected samples. In addition, to derive the
LF at  $z\ge1$ in a wider but shallow sample (say $K<18.5$)
will be essential to trace the evolution of the most massive galaxies 
with higher statistical significance. 
In the near future direct measurements of rest-frame $K$-band magnitude for
high redshift galaxies will be possible with space-based infrared surveys,
in particular  the SIRTF (Space Infrared Telescope Facility) Legacy Science 
project GOODS (Great Observatories Origins Deep Survey, P.I. M.
Dickinson).
Our data allow to directly estimate the number densities 
of $z<2$ galaxies at the different 
SIRTF-GOODS wavelengths,
$\lambda=3.6,4.5,5.8 \mu$m, which sample the
rest-frame $K_s$-band at $z\simeq0.7,1.1,1.7$ respectively, similar to
the $z_{mean}$ adopted for the K20 LFs. 
We have converted the absolute $M_{K_s}$ magnitudes, derived here for the K20 
sample,
to SIRTF+IRAC fluxes using the galaxy spectral models described in Sect. 
\ref{sec:LIR} for normal galaxies and the spectral model for M82 
(Silva et al. 1998) for the dusty star forming galaxies.
The expected number densities for objects with $z<2$ are shown in 
Fig. \ref{fig:MIRz}. 
At the expected depth of GOODS (see Dickinson et al. 2002b), 
SIRTF will detect about 5000 galaxies at 
$z<2$ within 330 arcmin$^2$.
GOODS SIRTF Legacy Program, sampling  the rest frame near-IR 
luminosities directly, will allow to derive the $K$-band Luminosity Function 
along with the stellar mass distribution
well below the K20 limit (see Fig. \ref{fig:MIRz})
and with higher statistical significance.
With these data the stellar mass assembly history of galaxies will be 
measured over a wide range of redshifts and cosmic time.

We make publicy available the $1/V_{max}$ estimates at 
http://www.arcetri.astro.it/~k20/releases/index.html.

\begin{acknowledgements}
      Part of this work was supported by the MURST (Cofin 2000-2001), ASI
and by the
European Community Research and Training Network ``The physics of the
intergalactic medium".
We are in debt with Carlton Baugh 
for providing the HMM predictions. 
\end{acknowledgements}

\end{document}